\documentclass[11pt]{article}
\usepackage{geometry}                
\usepackage{graphicx}
\usepackage{amssymb}
\usepackage{epstopdf}
\usepackage{tikz}
\usepackage{subfigure}

\usetikzlibrary{positioning}
\definecolor{lavander}{cmyk}{0,0.48,0,0}
\definecolor{violet}{cmyk}{0.79,0.88,0,0}
\definecolor{burntorange}{cmyk}{0,0.52,1,0}

\def\lav{lavander!90}
\def\oran{orange!30}

\tikzstyle{peers}=[draw,circle,black,bottom color=black,
                  top color= black, text=black,minimum width=5pt]
\tikzstyle{superpeers}=[draw,circle,black, left color=white,
                       text=violet,minimum width=5pt]
\tikzstyle{legendsp}=[rectangle, draw, black, rounded corners,
                     thin,bottom color=\oran, top color=white,
                     text=black, minimum width=1.5cm]
\tikzstyle{legendp}=[rectangle, draw, black, rounded corners, thin,
                     bottom color=\lav, top color= white,
                     text= violet, minimum width= 1.5cm]
\tikzstyle{legendin}=[rectangle, draw, red, rounded corners, thin,
                     text= red, minimum width= 1.5cm]
\tikzstyle{legendlen}=[rectangle, draw, thick, blue, rounded corners, thin,
                     text= blue, minimum width= 1.5cm]
\tikzstyle{legend_general}=[rectangle, rounded corners, thin,
                           burntorange, fill= white, draw, text=violet,
                           minimum width=1.5cm, minimum height=0.4cm]
\begin{document}

%
%

\title{ Learning and coordinating in a multilayer network}

\author{ Hayd\'ee Lugo\thanks{
corresponding author: hlugo@ccee.ucm.es.}\\
{\small Department of Quantitative Economics. Universidad Complutense de Madrid, 28223 Madrid, Spain} \\ \\
%
Maxi San Miguel\\
{\small IFISC (CSIC-UIB), Campus Universitat de les Illes Balears, 07122 Palma de Mallorca, Spain }}

\date{}                                           

\maketitle


\begin{abstract}
We introduce a two layer network model for social coordination incorporating two relevant ingredients: a) different networks of interaction to learn and to obtain a pay-off, and b) decision making processes based both on social and strategic motivations. Two populations of agents are distributed in two layers with intralayer learning processes and playing interlayer a coordination game. We find that the skepticism about the {\it wisdom of crowd} and the local connectivity are the driving forces  to accomplish full coordination of the two populations, while polarized coordinated layers are only possible for all-to-all interactions. Local interactions also allow for full coordination in the socially efficient Pareto-dominant strategy in spite of being the riskier one. 
\end{abstract}

\section{Introduction}

Several mechanisms and models have been implemented to explain the collective social behavior that arises from the interactions among individuals.  The own experience and the experiences of others play an important role in determining the people choices in almost all  human interactions. Imitation has been a widespread mechanism of human decision-making. Imitation of of {\it common} behavior reflects social influence in the individual, while imitation by others of a successful individual is of strategic nature [1,2,3,4]. Strategic interactions are often modeled by Game Theory. A relevant game theoretical model that describes  many  real-life interactions in which the best course of action is to conform to a consensus  is the coordination game. The challenge of such model is how to coordinate among its multiple Nash equilibria [5]. This issue has been addressed in several works focusing on coordination games in a network framework [6,7,8,9].
However, two relevant aspects of this context have been largely unexplored.

First,  the study of a kind of interactions in which individuals  distinguish according to their roles between people with whom they play to obtain a payoff and those from whom they learn to update their strategies.
An appropriate framework is needed to deal with the possibility that people may identify the kind of interaction they have with their partners. Such situations are very common and pertinent in real-life interactions. For example, the interactions between and within firms and consumers, employers and employees, governments and citizens, teachers and students, parents and children, medical doctors and patients. There individuals interact  across groups and receive a payoff for such interactions (for instance parents with children) and  look inside their group to learn and update their strategies (for instance parents learn from other parents and children learn from other children). What we have  are the situations in which  two populations  are differentiated by the role  that their individuals perform.
In simple models of social networks  individuals are unable to encompass different types of relationships. They play with and learn from the same set of neighbors. A  different class of networks  that have {\it layers} in addition to nodes and links, has been growing in popularity because of being a better
description of a real networked society.
The study and analysis of multilayer networks  is relatively recent  even though layered systems were examined decades ago in disciplines like sociology and engineering [10,11,12], for a complete review see [13]. Here we propose a two-layer network in which inside each layer, individuals update their strategies by a rule of learning and  across layers individuals  receive an aggregate payoff by playing a coordination game.
Most  previous studies of games in multilayer networks [14,15,16,17] consider playing the game inside the layers while we consider game theory interactions across layers. In a recent work  [18]  the authors consider a two-layer network wherein one network layer is used for the accumulation of payoffs playing a social dilemma game and the other is used for strategy updating.  There, each agent is simultaneously located on both layers. In contrast, in our  two-layer network, each agent is located in just one layer. Therefore, there are two learning networks, one in each layer, and a playing network across the two layers.

The second aspect refers to elucidate what happens when people make decisions heeding simultaneously social and strategic motivations, [4].
In situations that  call for accomplishing social efficiency and consensus two forces influence agent's choices:  the strategic reasoning and the social pressure of the environment.
In the sociological context, Granovetter [19]  proposed a model in which a certain amount of social pressure is necessary for a person to adopt a new idea, product or technology. Opinion, innovation spreading and social learning models have been dealing with this issue measuring the social pressure as the number of contacts that have already adopted the newness, [19,20,21,22].  Here, we consider that the influence of social pressure is related with the degree of doubts about the strategies currently being played.
Traditionally,  the degree of  doubts is measured as the subjective belief about the consequences of a certain action, [23].   However, we assume doubts as a social factor influencing choices in strategic environments. Then, the doubts of an agent about how well she is playing  depend on the popularity of her current strategy in her learning network.
Our approach of doubts is inspired by the work of  [24]. They introduce an evolutionary model of doubt-based selection dynamics. As well as [24], we assume that the agents measure their doubts by observing the choices made
by their fellow agents.  Real-life interactions and laboratory experiments [25,26,27] provide clear evidence of the importance of analyzing evolutionary dynamics based on social and strategic factors. For instance, in [4,28] the authors
explore the interplay between strategic and social
imitative behaviors  in a coordination problem on a social network and in a networked Prisoners' Dilemma respectively.
In these works agents can evolve by a mixed dynamics of the voter model [22,29] and the unconditional imitation. One of the main results in coordination games on complex networks is that the interplay of social and strategic imitation
drives the system towards global consensus while neither social or strategic imitation alone does. Our approach aims to deal with these two important aspects mentioned above and verify the circumstances in which the complexity of such social and strategic behavior leads to the consensus on the whole society.

The paper is organized as follows: in Sec. 2, we describe in detail our model of multilayer network  and  propose an update rule based on social and strategic factors. In section 3 and 4 we show and analyze  the numerical results for a pure and a general coordination game respectively. Finally, Sec. 5 summarizes our main findings.

\section{Model description}
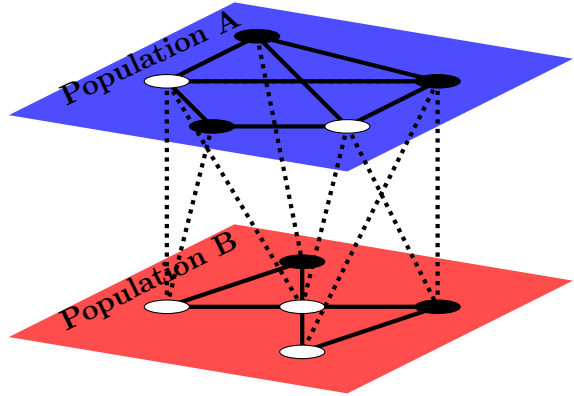
\begin{figure}
\begin{center}
\begin{tikzpicture}[scale=.6,every node/.style={minimum size=1cm},on grid]
		
    \begin{scope}[
    	yshift=0,every node/.append style={
    	    yslant=1,xslant=-2},yslant=0.5,xslant=-1.5
    	  ]
        \fill[blue,fill opacity=0.7] (0,0) rectangle (5,5);
    \end{scope}
    \begin{scope}[
        yshift=-140,every node/.append style={
        yslant=1,xslant=-2},yslant=0.5,xslant=-1.5
                  ]
            \fill[red,fill opacity=0.7] (0,0) rectangle (5,5);
    \end{scope} 

	\draw[color=black, dotted,ultra thick](-3,1) to (-4,-3);
	\draw[color=black,dotted,ultra thick](-2,3) to (-1,-2);		
	\draw[color=black,dotted,ultra thick](0,1) to (-1,-3);	
	\draw[color=black,dotted,ultra thick](-1,-3) to (-4,2);	
	\draw[color=black,dotted,ultra thick](2,2) to (-1,-4);	
	\draw[color=black,dotted,ultra thick](2,2) to (-4,2);	
	\draw[color=black,dotted,ultra thick](2,2) to (2,-3);	
	\draw[color=black,dotted, ultra thick](-4,2) to (-4,-3);
		\draw[color=black,dotted,ultra thick](0,1) to (2,-3);
		
	\draw[color=black,ultra thick](-3,1) to (0,1);
	\draw[color=black,ultra thick](-3,1) to (-4,2);		
	\draw[color=black,ultra thick](0,1) to (2,2);	
	\draw[color=black,ultra thick](-2,3) to (2,2);
	\draw[color=black,ultra thick](-2,3) to (0,1);
	\draw[color=black,ultra thick](-4,2) to (-2,3);
	\draw[color=black,ultra thick](2,2) to (-4,2);	
	
	\draw[color=black,ultra thick](-1,-2) to (-1,-3);	
	\draw[color=black,ultra thick](-4,-3) to (-1,-3);	
	\draw[color=black,ultra thick](2,-3) to (-1,-3);	
	\draw[color=black,ultra thick](-1,-4) to (-1,-3);	
	\draw[color=black,ultra thick](-4,-3) to (-1,-2);	
	\draw[color=black,ultra thick](-1,-4) to (2,-3);		
	
	\draw[fill=black]
			(-3,1) ellipse (0.5cm and 0.15cm) 
			(-2,3) ellipse (0.5cm and 0.15cm) 
			(2,2) ellipse (0.5cm and 0.15cm) 
			(-1,-2) ellipse (0.5cm and 0.15cm) 
			(2,-3) ellipse (0.5cm and 0.15cm); 
\draw[fill=white]
			(-4,2) ellipse (0.5cm and 0.15cm) 
			(0,1) ellipse (0.5cm and 0.15cm) 
			(-1,-4) ellipse (0.5cm and 0.15cm) 
			(-4,-3) ellipse (0.5cm and 0.15cm) 
			(-1,-3) ellipse (0.5cm and 0.15cm); 
			
			\fill[black]
			(-6.5,1.5) node[right, scale=1,rotate=25] {\textbf{Population A}}
			(-6.5,-3.5) node[right,scale=1,rotate=26]{\textbf{Population B}};
\end{tikzpicture}
\end{center}
\caption{Sketch of a multilayer network that we consider in this paper. The nodes are connected to each other in a pairwise manner both inside of the layers and between the layers for two populations A and B. Dotted lines describe the playing network (i.e. interlayer edges) and the solid lines describe the learning network (intralayer edges). The black nodes describe the agents playing strategy L and white nodes the agents playing strategy R in a  coordination game.}
\label{multilayer}
\end{figure}

 In this paper we consider a two-layer network in which each individual is connected to two different social networks, the interlayer network or playing network, and the intralayer network or  learning network, see Figure \ref{multilayer}.
 \subsection{Coordination games}
 In the playing network, each  player interacts according to a coordination game with each of her neighbors using the same action for all those games.
A normal form representation of this two-person, two-strategy coordination game is shown in Table \ref{gcg}. We focus our analysis in two parametric settings, a pure or symmetric coordination game in which $a=d=1$ and $b=0$ and a general or asymmetric coordination game in which $a=1$, $d=2$ and $b>0$.
The profiles $(L,L)$ and $(R,R)$ are the two Nash equilibria in pure strategies in both settings. Now, in the general coordination game the agents get a higher payoff by playing  $(R,R)$,  the Pareto (payoff) dominant equilibrium while for $b >1$ they risk less by coordinating on $(L,L)$, called  the risk dominant equilibrium.  Games of this type are more interesting than their fully symmetrical versions as it is added a confidence problem when the socially efficient solution is also the riskier one.

\begin{table}[htdp]
\begin{center}
\begin{tikzpicture}[scale=1.3]
\draw (1,0) grid (3,2);
\node at (0.5,1.5) {$ L$};
\node at (0.5,0.5) {$R$};
\node at (1.5,1.5) {$a, a$};
\node at (2.5,1.5) {$0, -b$};
\node at (1.5,0.5) {$-b, 0$};
\node at (2.5,0.5) {$d, d$};
\node at (1.5,2.2) {$L$};
\node at (2.5,2.2) {$R$};
\end{tikzpicture}
\end{center}
\caption{Pay-off matrix for a general two-person, two-strategy coordination game}
\label{gcg}
\end{table}%
\subsection{Doubts and the parameter T}
 In the learning network, we propose an evolutionary update rule that heeds strategic thinking and the doubts that are generated by the popularity of the strategies.
  In order to describe this aspect in detail we provide some definitions.
As  Cabrales and Uriarte [24], we assume that the doubts  felt by an agent are related to the proportion of individuals with whom they interact who are equally using
the same strategy.
Our approach differs from [24]  since while  those authors  assume that the agents are endowed with a {\it
doubt function}, we assume that they are endowed with a quantity $T$  that calibrates their level of doubts about  the collective {\it wisdom of crowd},  $T \in[0,1]$.  This parameter $T$ is in the same line of the threshold value in [19]. Just as in [24], we may distinguish two broad types of population, each corresponding to a doubtful behavior. {\it A herding population}, for $T<0.5$, is a population in which agents rely on the {\it wisdom of crowd}. As a consequence, they are strongly  influenced by the popularity of the current strategies of their partners. {\it A skeptical population},  for $T \ge 0.5$, is a population in which agents  are very suspicious of the {\it wisdom of crowd}: they are slightly  influenced by the popularity of  the current strategies of their partners.
In the updating process, each player $i$  observes the proportion of agents,  $d_i $,  who are playing the opposite strategy to hers in her learning neighborhood.
Then, she measures how popular her strategy is, comparing $d_i$ with $T$. For instance, when $d_i > T$, player $i$ has  doubts about the popularity of the strategy she is currently playing.

\subsection{The degree of dissatisfaction}
The evolution on time of the strategies derives from the levels of dissatisfaction felt by the agents.
The criterion that defines the level of satisfaction of an agent is based on two key points: how well she is doing in terms of the payoff  obtained in her playing network  and how popular  her current strategy is in her learning network.
Our approach of satisfaction is quite different from [24] where they justify  the  choice of
an {\it index of dissatisfied agent}   via a model of  {\it (correlated) similarities relations}  and from [9] that define a quantity called satisfaction based on the strengths of the links.
In our approach we distinguish four categories of agents as described in Table \ref{sa}, where $\pi_i$ is the aggregate payoff of agent $i$ and $n_i$ is  her degree in the playing network.:
\begin{table}[htdp]
\begin{center}
\begin{tabular}{c|c|c|c|c}
& S  &  P1 & P2  &	U\\ \hline
playing network & $\pi_i=\beta n_i$  & $\pi_i=\beta n_i$ & $\pi_i <\beta  n_i$  &$\pi_i< \beta n_i$	\\
learning network &$d_i < T$ &  $d_i > T$ & $d_i < T$ & $d_i > T$ \\  \hline
\end{tabular}
\caption{Degrees of satisfaction}
\label{sa}
\end{center}
\end{table}%

\noindent   The value of $\beta$ is derived from the parametric setting of the class of coordination game played.
Since in the pairwise interaction of pure coordination games each player gets a payoff of 1 by coordinating  and 0 otherwise, then $\beta=1$ for such game.
The equality  $\pi_i=n_i$ means that  the player $i$ coordinates with all her neighbors in the playing network: then we say  that agent $i$ is strategically satisfied. In the case of a general coordination game, $\beta = 2$ and an agent  is strategically unsatisfied when she fails to coordinate with all her neighbors on the socially efficient solution, {\it i.e.} the Pareto dominant strategy. This will happen when in a time step  $\pi_i < 2n_i$.
When $d_i < T$ the proportion
of neighbors in her learning network who play the same strategy as she does is high enough so that player $i$ feels socially satisfied with her current strategy. Then,
the level of satisfaction of an agent $i$ is: S (satisfied) when she is both socially ($d_i < T$) and strategically ($\pi_i=\beta n_i$) satisfied, is
P1 or P2 (partially satisfied) when she is either socially ($d_i >T$) or strategically ($\pi_i < \beta  n_i$) unsatisfied and is
 U (unsatisfied) when she is both socially ($d_i >T$) and strategically ($\pi_i <\beta  n_i$) unsatisfied.

\subsection{The strategic update rule}
We propose a synchronous update rule in which  each player can change her current strategy according to her level of satisfaction. Namely,
\begin{enumerate}
\item If her level of satisfaction is S,  she remains with the same strategy.
\item If her level of satisfaction is P1 or P2,   she imitates the strategy of her best performing neighbor in her learning network when such neighbor has received a larger payoff  than the player herself, otherwise she remains with the same strategy. 
\item If her level of satisfaction is U, she changes her current strategy.
\end{enumerate}
This rule might resemble  the well-known unconditional imitation (UI) update rule introduced in [30]. When agents follow the (UI) update rule, they seek to maximize their payoffs imitating the most successful individuals.   However, the first important difference in our update rule is that individuals change their strategies conditional to their social or strategic dissatisfaction.  Some experimental results show evidence of the use of the (UI) rule by individuals but also provide evidence that other social factors are influencing the updating process [25,26,27].
Other important difference is the environment in which learning takes place. Since individuals discriminate from whom they learn and with whom they play, this update rule only takes place in the learning networks. The proposed update rule aims to capture the individual behavior in a complex real life situation.
Having setting out our strategic and social framework, we now turn to describe the evolutionary dynamics.
At each elementary time step, each player plays the coordination game with each one of her {\it interlayer} neighbors. Once the game is over and a payoff is assigned to each player, each agent, observing her {\it intralayer} neighbor,s might change her strategy according to her level of dissatisfaction. The process is repeated setting payoffs to zero.

\subsection{Simulation settings}
The  size of the populations  A and B  during simulations is $N_A=N_B=1000$.
The numerical results are obtained for random (Erd\"{o}s- R\'enyi, ER) networks and fully connected networks.  In the learning networks, $k_{AA}$ and $k_{BB}$ represent the mean degree (average number of links per node)  for population $A$ and $B$ respectively. In the playing network,  the mean degree $k_{AB}$ corresponds to the average number of links per node across populations $A$ and $B$.
 The two strategies of the coordination games are  L and R which are initially uniformly randomly distributed with proportion $0.5$.

\section{Results for pure coordination games}
As a benchmark,  it is helpful to remind the final configuration of a structured population playing a pure coordination game with the  (UI) as update rule. The topology will define the outcome of such population. For instance, for a complete network, referred also as a fully connected network,  in which each agent interact with every other agent, full coordination is reached in one time step, while for a social network displaying local connectivity, such as the random (ER) network,  the system evolves to a non-coordinated frozen state.
For the study of our model we focus on these two network topologies. Our simulation results  show that the combination of strategic and social factors in a multilayer network  drives the
system to quite different outcomes than those ones.
Before displaying the results, we need to clarify what  a complete network means  in our context of multilayer network. A complete network here implies that every agent plays with every other agent in the playing network and learns from every other agent in the learning network.  Agents still discriminate between with whom they play and from whom they learn.
Moreover an absorbing state in this framework is
a state of intralayer coordination. In this state the agents are socially satisfied since inside each layer the same strategy is spreading all over the network. A state of interlayer coordination is a state of intralayer coordination in which the strategy displayed in one layer coincides with the strategy reached in the other layer: agents are socially and strategically satisfied. However, when the strategy in one layer is  the opposite to the one in the other layer, the social satisfaction of agents makes the strategies to remain unchanged, and the configuration of a polarized two-layer network is an absorbing state of the dynamics. In summary, a state of interlayer coordination implies a  state of intralayer coordination.  but the reciprocal is not necessarily fulfilled. Both interlayer coordination (or full coordination) and intralayer coordination are absorbing states of the dynamics. 

\begin{figure}[]
\begin{center}
\includegraphics[width=.47\textwidth]{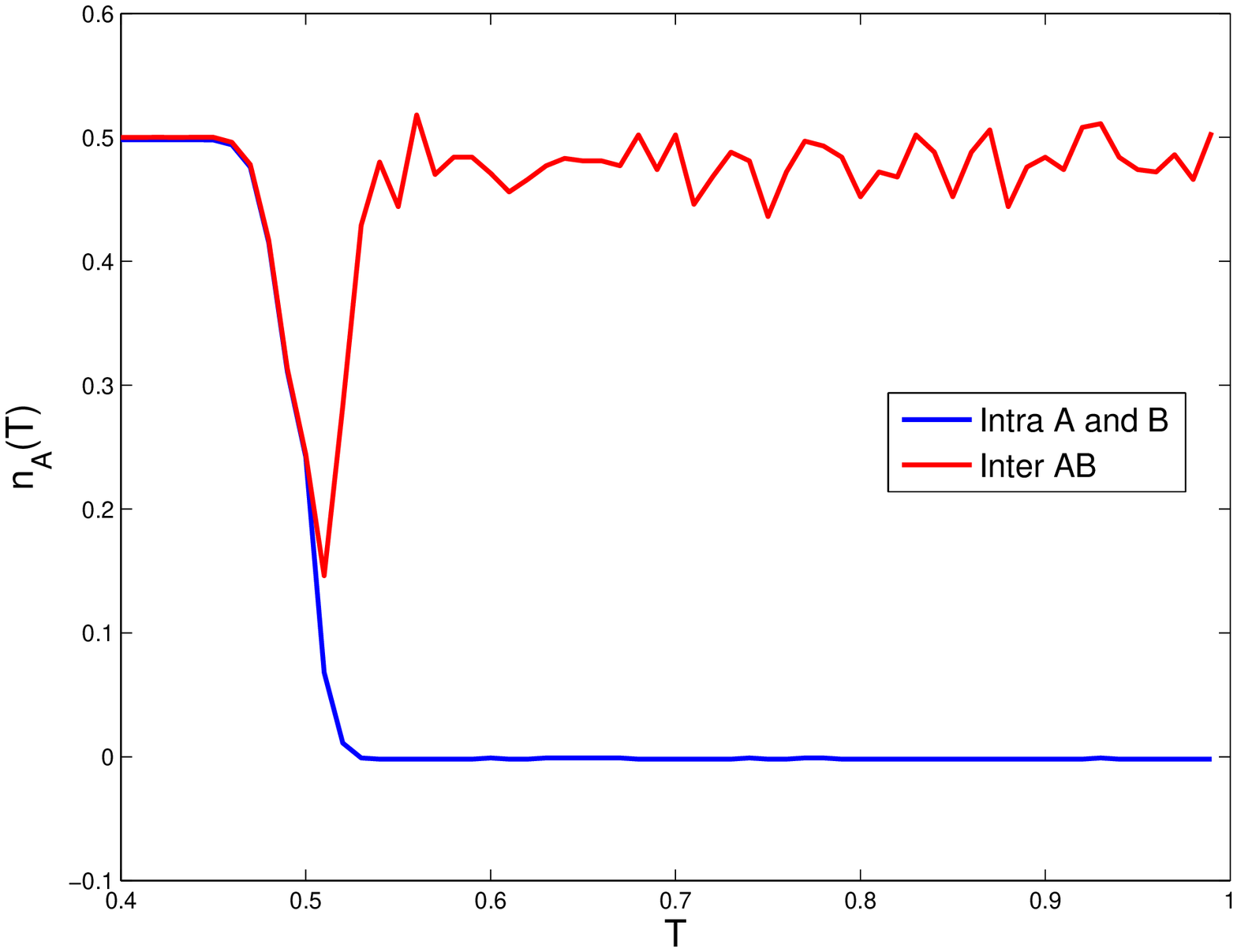}
\includegraphics[width=.45\textwidth]{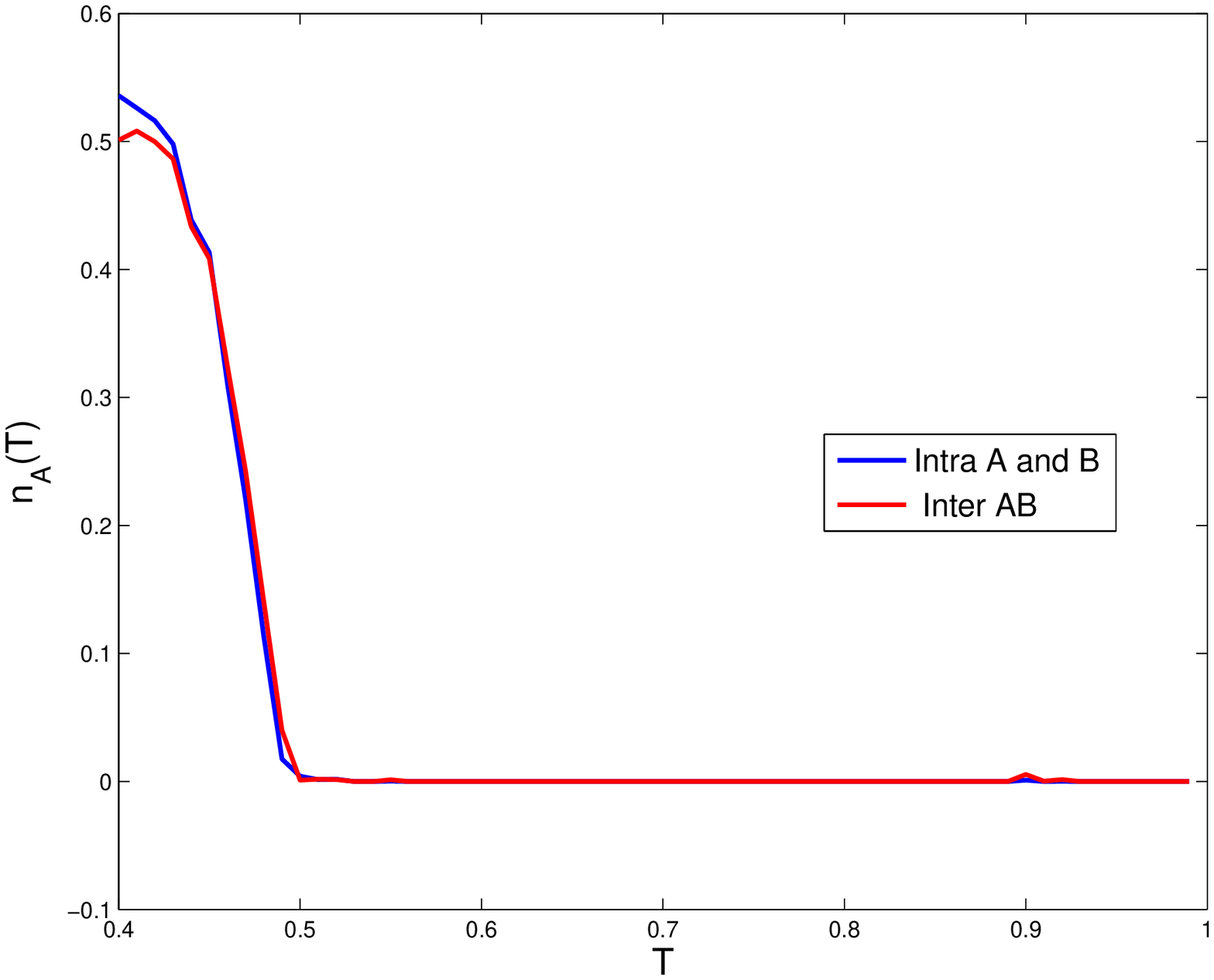}
\caption{Average over 500 realizations of the densities of intralayer active links for each population A and B  and interlayer active links between A and B  for each $T \in [0.4,1]$ with $N_A=N_B=1000$  in a fully connected network (left panel) and in a random (ER) network with $k_{AA}=k_{BB}=k_{AB} = 10$ (right panel).}
\label{alpcg}
\end{center}
\end{figure}

 The final configurations of the system can be described by the intra (inter) active links defined as the number of links connecting agents with different choices  in the playing (learning) network.
Figure \ref{alpcg} shows the average of the proportions of  active links $n_A(T)$,   inter layers between populations A and B and intra layer for each population, A and B, for  $T \in [0.4,1]$  in  the fully connected network (left panel)  and in the random (ER) network  (right panel). We find that
for herding populations, $T < 0.5$, the final configuration of the system is a state of non-coordination in both the learning network (intralayer) and  the playing network (interlayer) for the fully connected network and the random (ER) network. Too much sensitivity to the social pressure plays against the intralayer, and therefore, the interlayer coordination in any of these two network topologies. Such non-coordination state is the one in which the proportions of the strategies in population A and B fluctuate over $0.5$, see the left panel of Figure \ref{ss}.
However, in the case of skeptical populations, $T \ge 0.5$, the system always reaches intralayer coordination both in the fully connected and in the random (ER) networks. However, for interlayer coordinations, we observe coordination on all realizations of the process in the case of random (ER) networks, while interlayer coordination is only reached in half of the realizations in the fully connected network.
Figure \ref{coorpcg} shows  the number of realizations in which the system reaches a state of interlayer coordination on the strategy L and R and a interlayer non-coordination state for  $T \in [0.4,1]$. For $T >0.5$, we observe that in the fully connected network (left panel), agents fully coordinate either in L or R half of the realizations. The steady state of non interlayer coordination is a completely polarized multilayer network in which all agents in population A play the opposite strategy of all agents in B, see the right panel of Figure \ref{ss}. In the case of  random (ER)
networks (right panel of Fig.3) a state of interlayer coordination either in L or R is always reached for $T >0.5$. Comparison of this result with the one for fully connected networks highlights the role of local interactions to reach consensus or full coordination: While with all-to-all interactions (fully connected networks) interlayer coordination is only reached in half of the realizations, the presence of local interactions (ER networks) leads always to full (interlayer) coordination for skeptical populations ($T> 0.5$).

\begin{figure}
\label{coorpcg}
\begin{center}
\includegraphics[width=.4\textwidth]{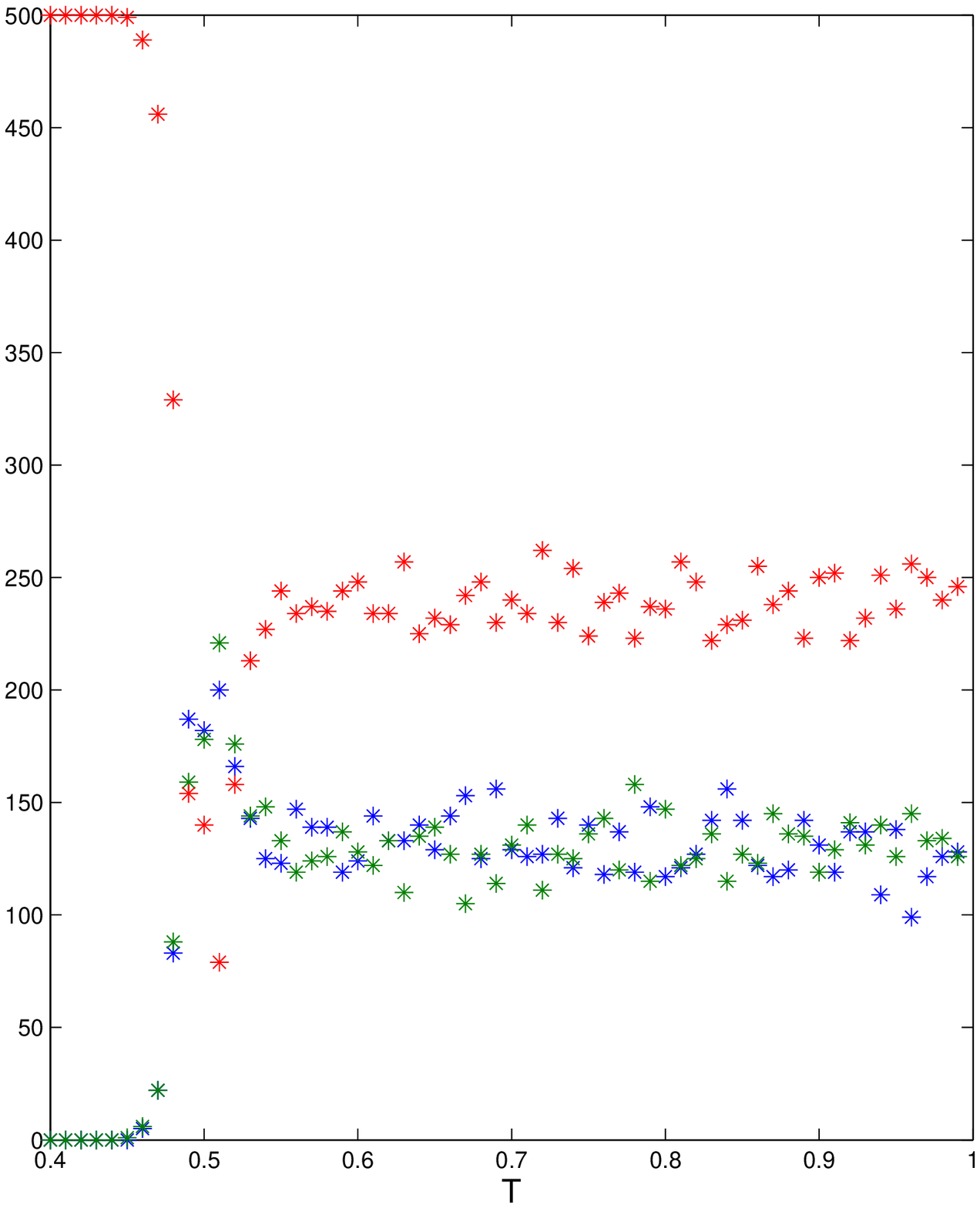}
\includegraphics[width=.4\textwidth]{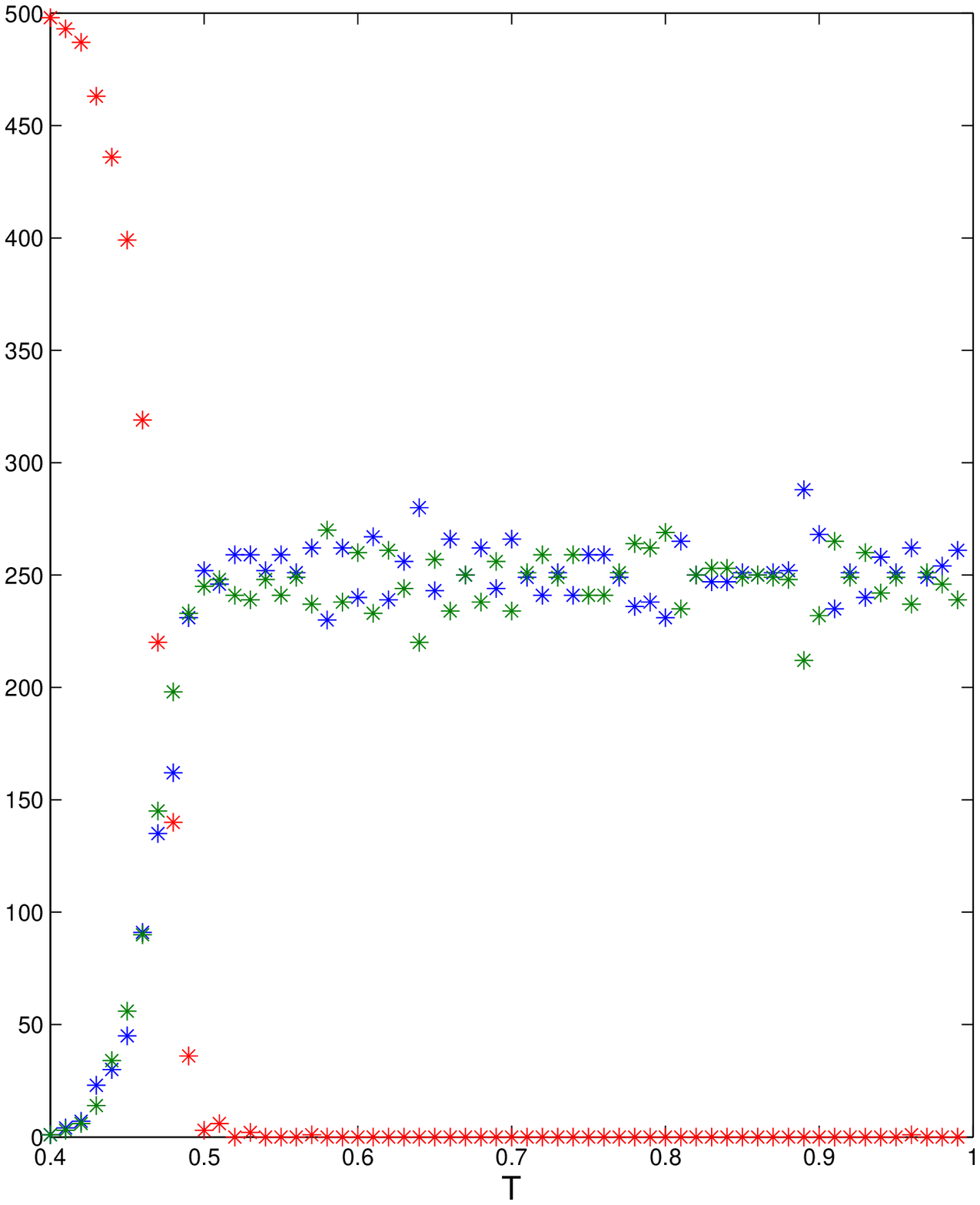}
\caption{Number of realizations, out of a total of 500 realization, that the populations A and B reach coordination using strategy R (blue), using strategy L (green)  and are not able to coordinate (red) as function of T for a fully connected network(left panel) and $k_{AA}=k_{BB}=k_{AB}=10$ (right panel).}
\label{coorpcg}
\end{center}
\end{figure}

\begin{figure}
\begin{center}
\includegraphics[width=0.45\textwidth]{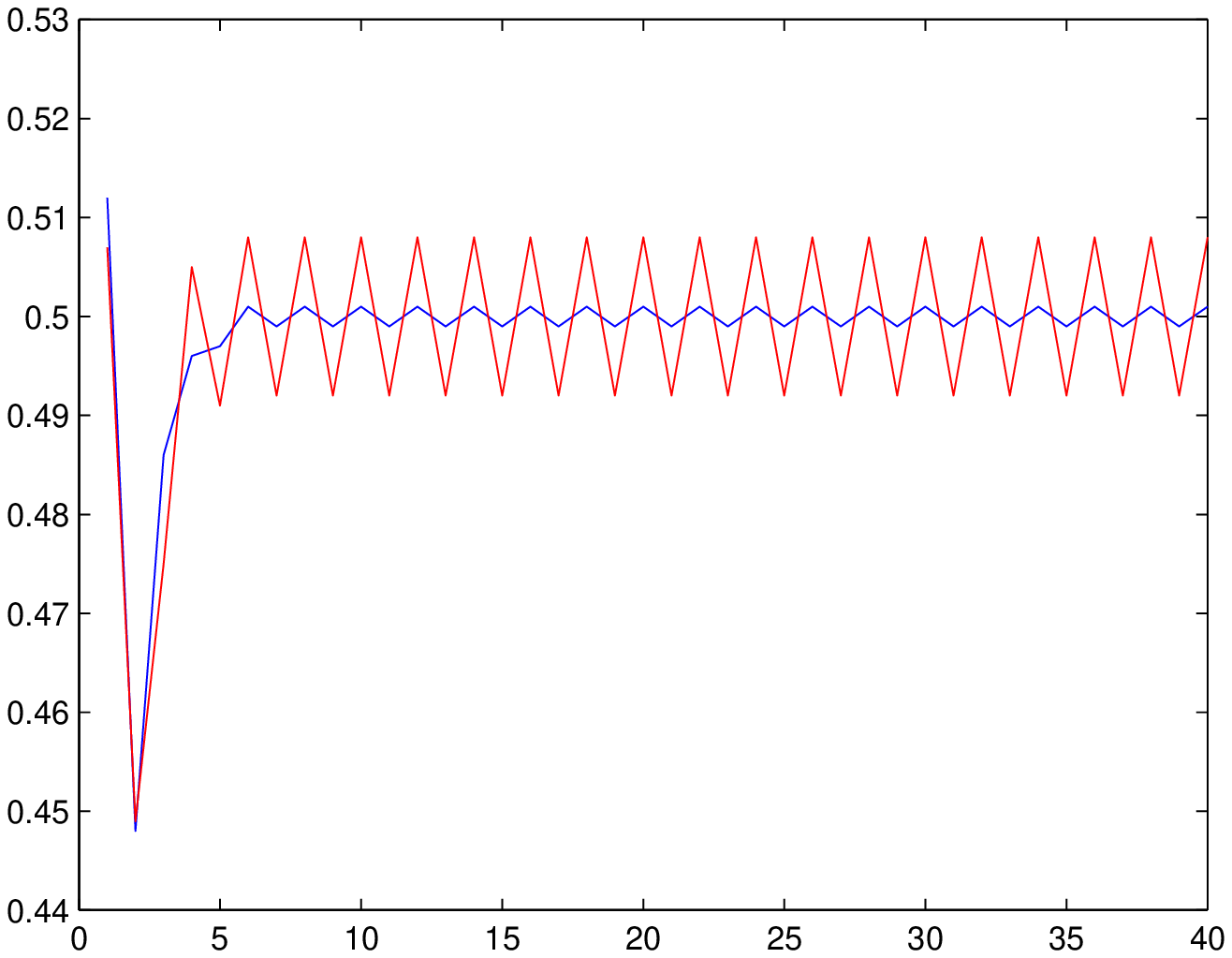}
\includegraphics[width=0.45\textwidth]{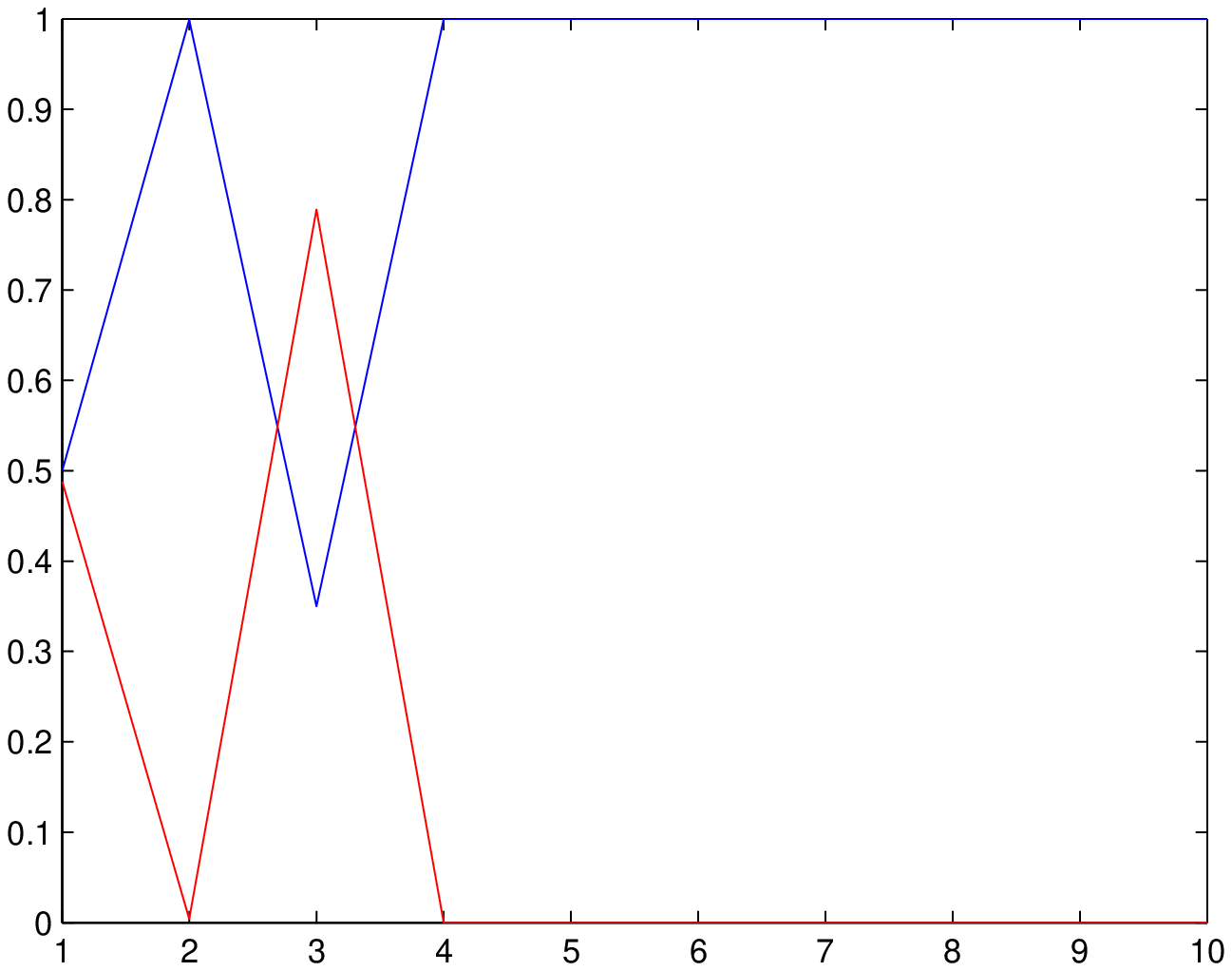}
\caption{Time series of the proportion of agents playing L in A (blue) and L in B (red)  in a random (ER) network with $k_{AA}=k_{BB}=k_{AB}=10$ and $T=0.3$ (left panel) and $T=0.8$ (right panel).}
\label{ss}
\end{center}
\end{figure}

\section{Results for general coordination games}
To cover a better understanding of this multilayer model,  we extend our analysis to a general coordination game setup whose normal form representation is shown in Table 1 with $a=1$, $d=2$ and $b>0$. Due to their social and strategic implications this class of games has been studied  analytically in an evolutionary framework  [31,32] and by  the numerical simulations on several network topologies [6,7,8].
Previous numerical results have shown that in a fully connected network, the agents using  the (UI) update rule tend to coordinate  on $(L,L)$, the risk dominant equilibrium whenever $b > 1$ and in the case of a complex network,  the (UI) update rule leads to frozen disordered configurations.
In our multilayer model with the dynamic update rule based on social and strategic implications,
our numerical results are again quite different from these previous results and also are determined by the doubtful behavior of the populations.
The same analysis made in the last section for pure coordination games leads too to the same conclusion that states of intralayer coordination are absorbing states of the dynamics. The state of interlayer coordination is another absorbing state that implies intralayer coordination.

\begin{figure}
\begin{center}
\includegraphics[width=.45\textwidth]{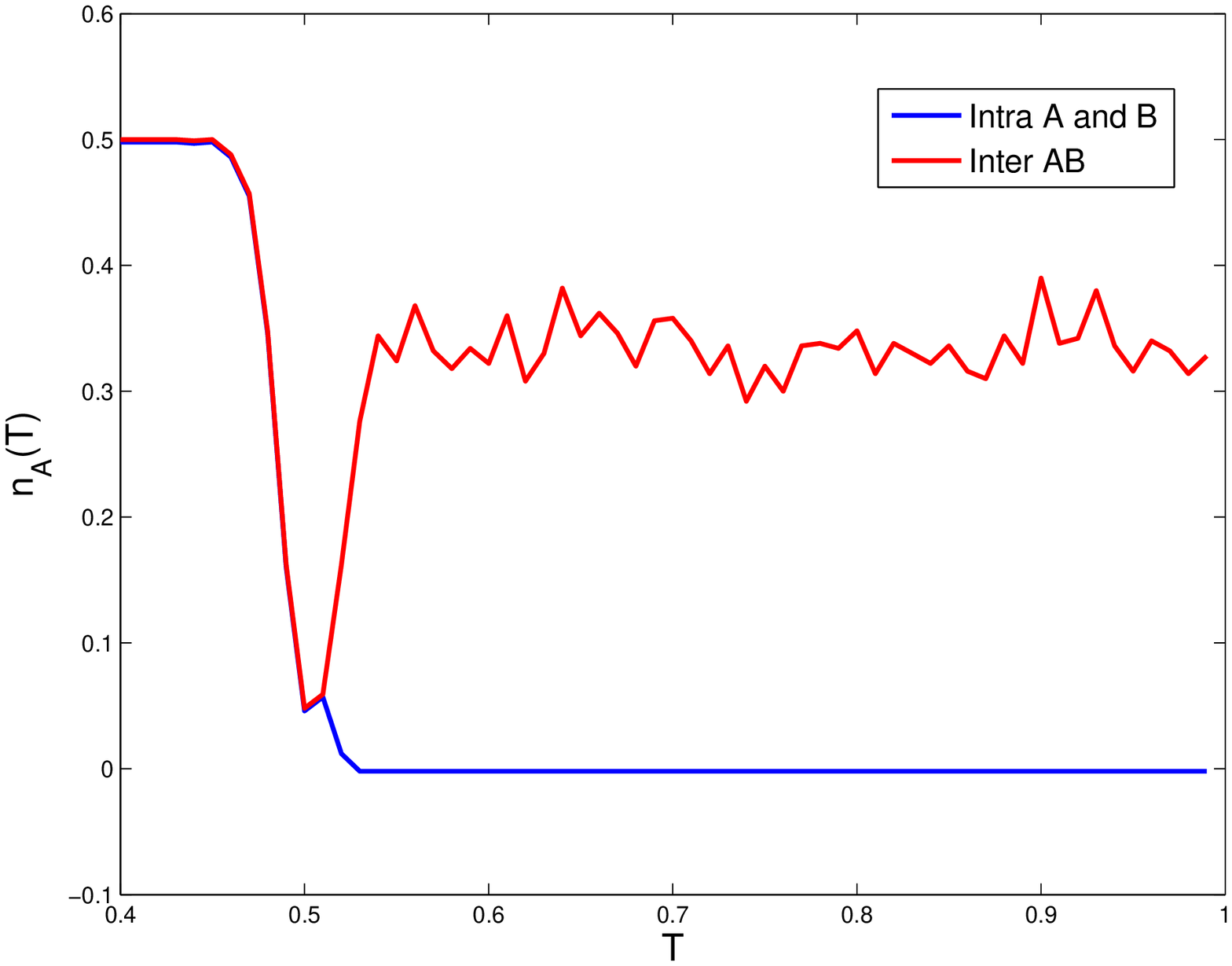}
\includegraphics[width=.44\textwidth]{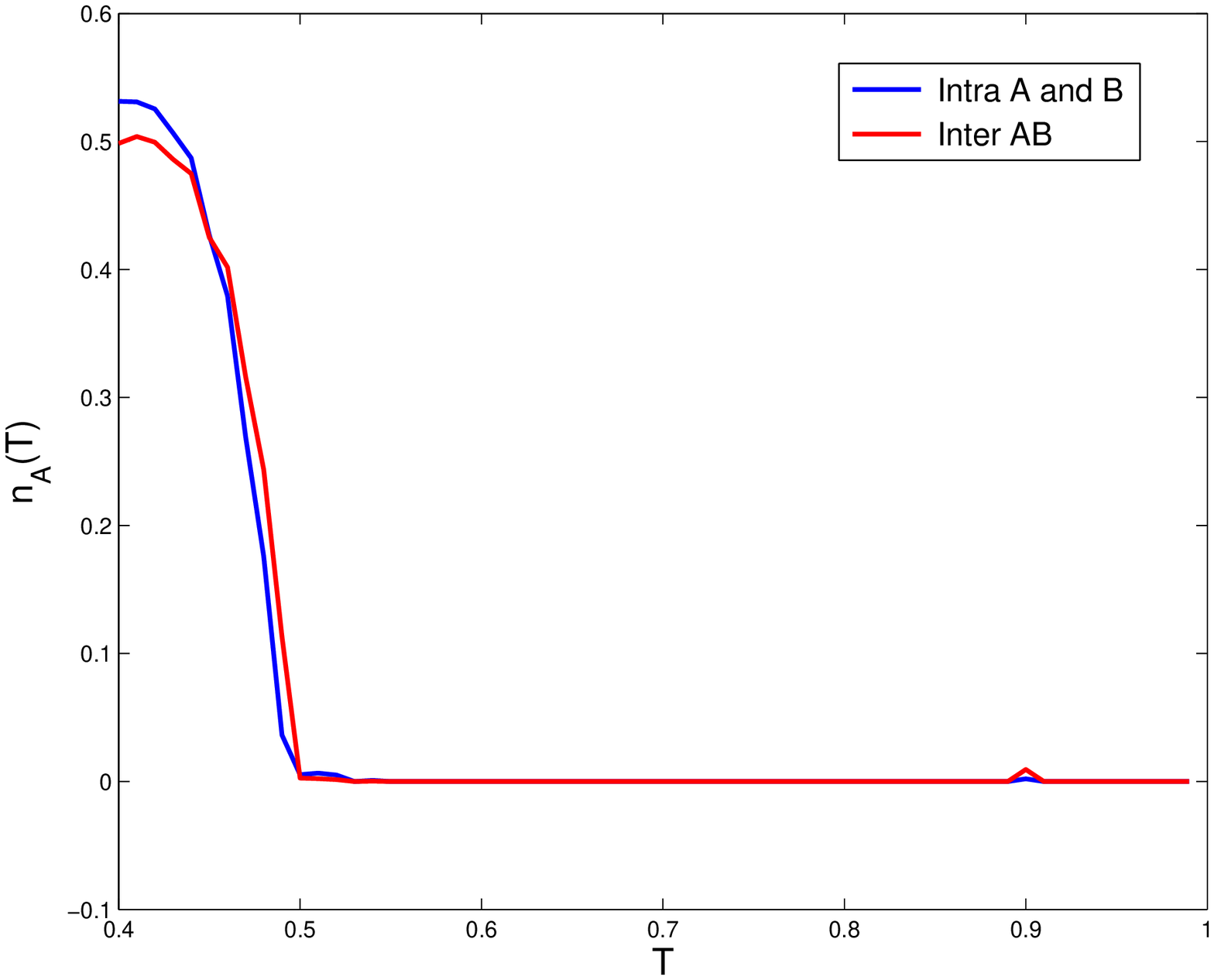}
\caption{For a general coordination game with $b=1.1$, average over 500 realizations of the densities of intralayer active links for each population A and B  and interlayer active links between A and B  for  $T \in [0.4,1]$ in a fully connected network (left panel) and a random (ER) network with  $k_{AA}=k_{BB}=k_{AB}=10$ (right panel).}
\label{algc}
\end{center}
\end{figure}

As already seen in the previous section of pure coordination game, also in the general coordination game the herding populations  are not able to reach intralayer coordination
 neither for fully connected nor for random (ER) networks. The sensitivity to the social pressure is again a detrimental factor in any of the two network topologies.
Similarly, for skeptical populations, the final configuration of intralayer coordination is always reached and depending on the network topology the state of  interlayer coordination is also accomplished.
As an example,  Figure \ref{algc} shows the densities of intralayer and interlayer active links for a general coordination game with $b=1.1$. For $T >0.5$, the system reaches  interlayer coordination almost $70\%$ of the realizations in the fully connected network ( left panel of Figure \ref{algc}). This proportion is higher than the $50\%$ observed in the case of the pure coordination games. In the random (ER) network, the final configuration of the system is always of interlayer coordination, see right panel of Figure  \ref{algc}.

\begin{figure}[h]
\centering
\subfigure[]{ \includegraphics[width=4.5cm,height=6.5cm]{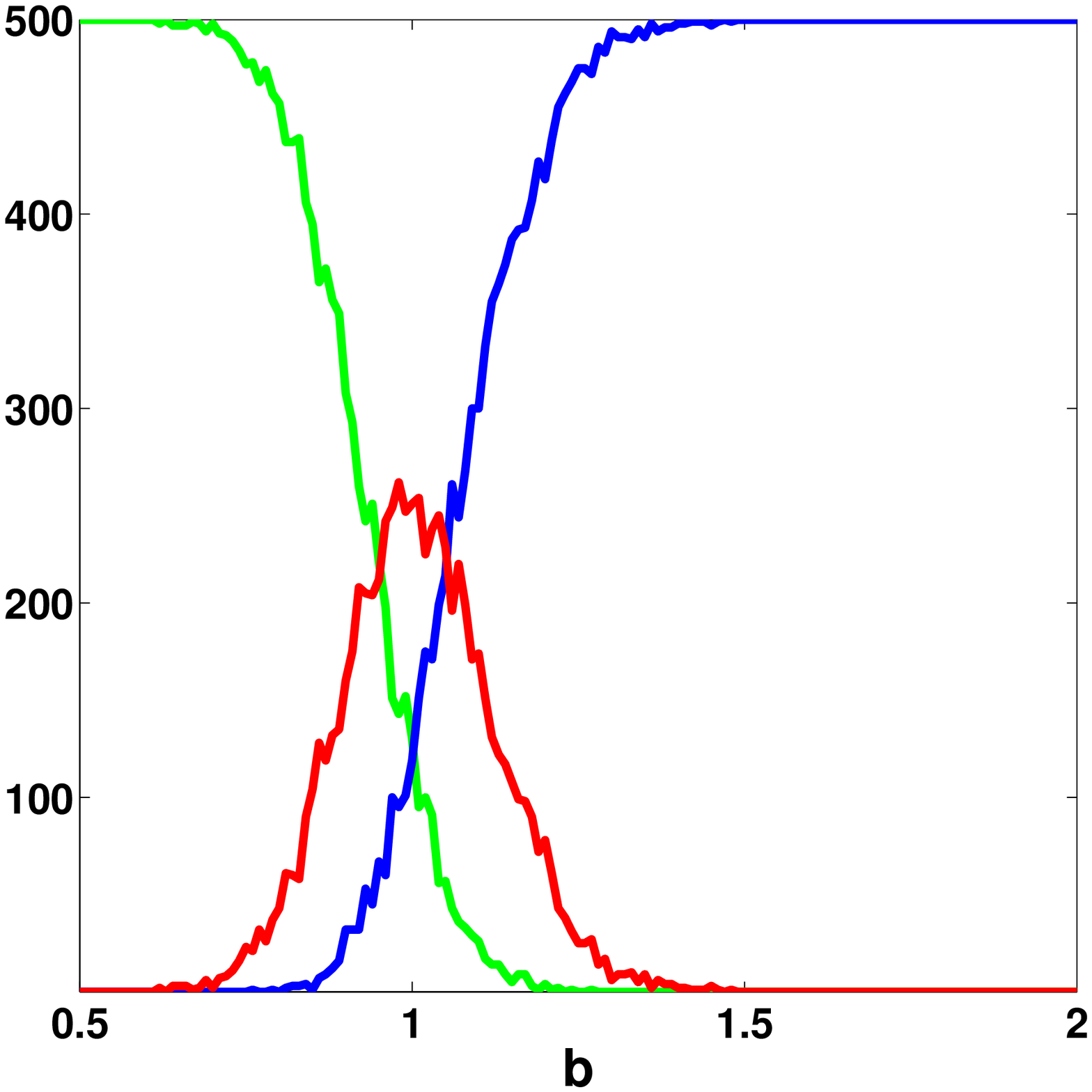} }
\subfigure[]{ \includegraphics[width=4.5cm,height=6.4cm]{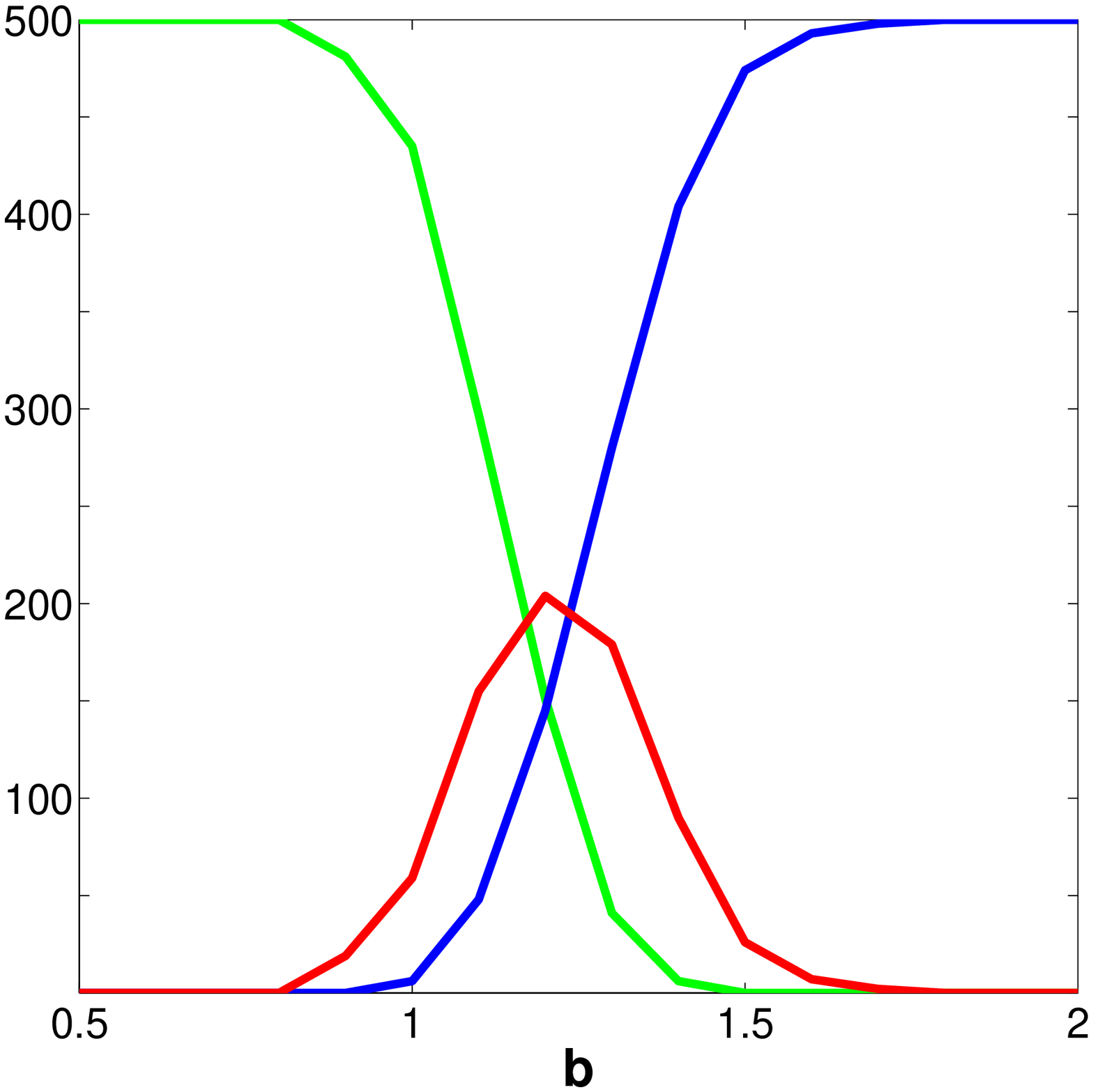} }
\subfigure[]{ \includegraphics[width=4.5cm,height=6.4cm]{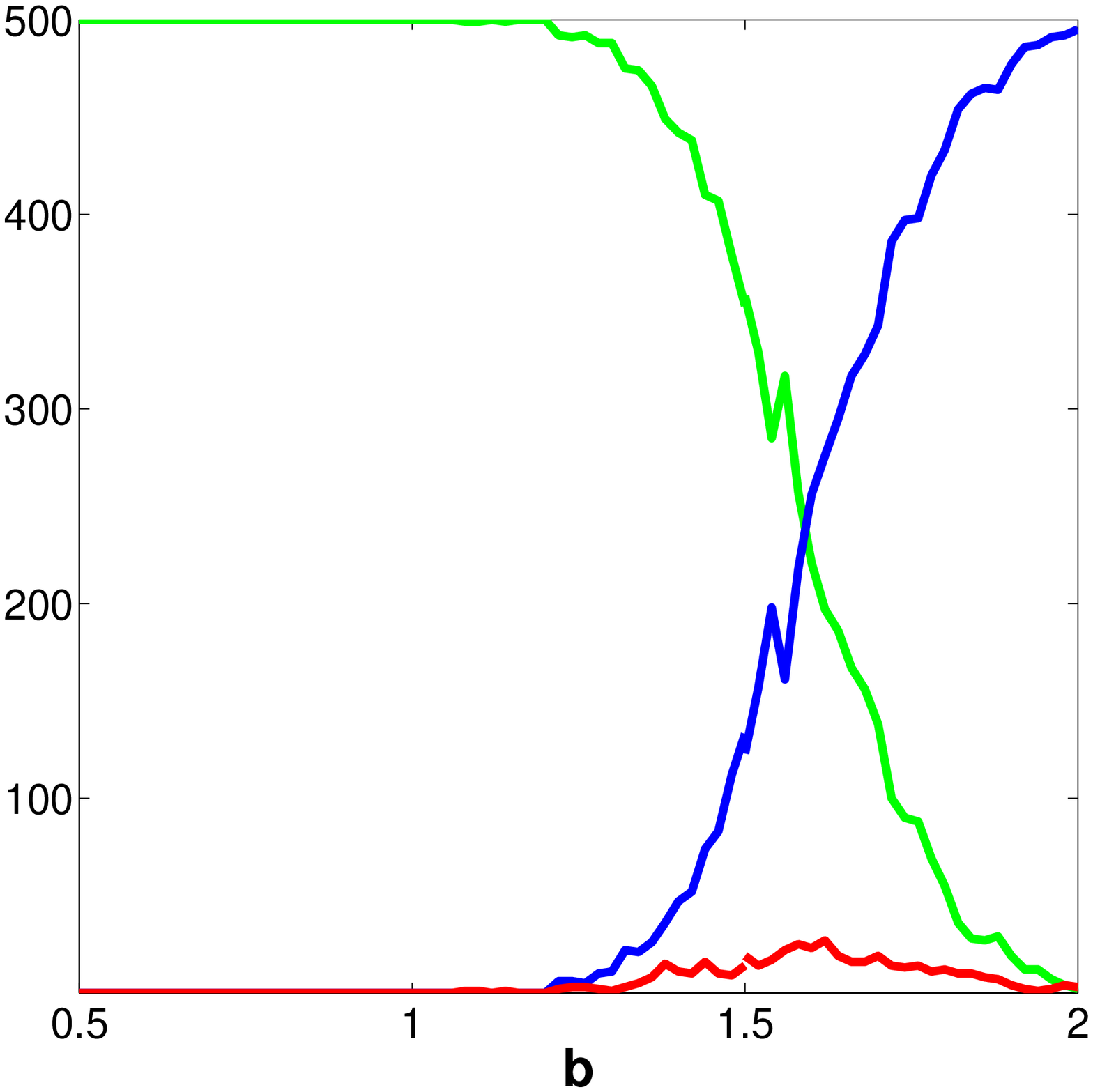} }
\caption{Number of realizations reaching interlayer coordination on L (blue),  on  R (green) and  polarized networks state with only intralayer coordination (red) as function of b for 500 realizations for skeptical populations with $T=0.7$, in a fully connected network (a),   a random (ER) network with $k_{AA}=k_{BB}=k_{AB}=500$ (b), and   $k_{AA}=k_{BB}=k_{AB}=100$ (c).}
\label{coorgcg}
\end{figure}

The main point at issue here is whether Pareto-dominant equilibrium can be coordinated by the agents. In the game theoretical approach, the coordination on the risk-dominant equilibrium  (L,L) is unavoidable whenever $b>1$. In our framework, skeptical individuals are those able to reach intralayer or interlayer coordination, however the key point is to find out whether such coordination favors the desirable socially efficient outcome, that is the (R,R) Pareto dominant coordination. First, let us analyze what happens in the complete multilayer network. As the initial strategies are uniformly randomly distributed with proportion $0.5$, almost all individuals are at least strategically unsatisfied and willing to change their strategies. According to the update rule, an unsatisfied agent who is playing $L$ in a fully connected network  will change her strategy to $R$ only when $b < \frac{2}{p_L}- 3$ where $p_L$ is the proportion of  agents playing $L$ in her learning network. Due to the initial conditions $p_L \approx  0.5$ the parameter $b$ must be approximately lesser than $1$ to make agents who are playing $L$ change to $R$.  Panel (a) of Figure \ref{coorgcg} shows, for a fully connected network, the number of realizations that the system reaches  interlayer coordination on $L$, on $R$, and intra but not interlayer coordination as function of $b$. We observe that as $b$ increases, the number of realizations reaching interlayer coordination on $L$ increases. As a consequence, the rate of coordination on the Pareto dominant equilibrium $(R,R)$ decreases with  $b$, with the most likely coordination shifting from Pareto dominance to risk dominance around $b*=1$, as expected. Noteworthy that the range of values of $b$ in which the state of polarized layers can be reached is also around $b=1$,  where the two Nash equilibria have the same expected payoff.
In  panels (b) and (c) of Figure \ref{coorgcg} for ER networks, we show that such  threshold $b^*$ in which  the chance of coordination on $R$ starts to decrease is higher the lower the average number of links  per node is. The effect of locality not only favors interlayer coordination over only intralayer coordination (polarized layers) but also favors Pareto dominant coordination.  In our numerical simulations (not shown) we find that already  for $k_{AA}=k_{BB}=k_{AB}=10$  the agents manage to coordinate on the Pareto dominant equilibrium $(R,R)$  for any value of $b \in [0.5,2]$, overcoming the frozen disordered configurations reported in previous works. The strong effect of locality is due to the possibility that $p_R > T$ for an agent who is playing $L$. In such case she will be totally unsatisfied and will switch her strategy to $R$.
 In our multilayer model, the locality for skeptical populations is the driving force that favors interlayer coordination on the socially efficient outcome, that is the Pareto dominant strategy.

\section{Conclusions}

In this paper we have introduced a multilayer network model in which agents of two populations play and learn in two disaggregated networks and  update their strategies heeding social and strategic motivations. A network between the two populations is for playing according to a  coordination game. There each agent receives an aggregate payoff as a result of her interaction with each of her playing neighbors. The other network is for learning in which each agent can update her game strategy motivated by a feeling of social or strategic dissatisfaction. When an agent is unsatisfied either socially or strategically, she can update her strategy imitating the strategy of the most successful neighbor. The agent searches for such neighbor looking inside her own population. We have shown that the degree of social pressure calibrated by the level of doubts plays an important role in the networks topologies considered.  The skepticism about the {\it wisdom of crowd} and the locality of interactions are the driving forces for collaboration and social efficiency in both pure and general coordination games.

For pure coordination games  in a skeptical environment, each population evolves towards a coordinated state in both fully connected and random (ER) networks. However, in fully connected networks (non-local interactions) the populations eventually may coordinate each other in the opposite strategy leading to a polarized multilayer network.
 In the case of general coordination games the challenge is to elucidate whether the Pareto-dominant strategy, the socially efficient
outcome, can be established in the populations. Previous results in well-mixed
and structured populations tend to favor the risk-dominant equilibrium in the parametric setting in which the Pareto dominant equilibrium is also the riskier one. In contrast,
our simulation results show that the skepticism and the local connectivity allow the populations  to coordinate on the Pareto-dominant equilibrium even in the riskier setting.

\section*{Acknowledgements}
H. Lugo gratefully acknowledges the financial support from Spanish MICINN under project ECO2013-42710-P.
M. San Miguel  gratefully acknowledges the financial support from the Spanish MINECO and FEDER under project INTENSE@COSYP (FIS2012-
30634). Authors  gratefully acknowledge the financial support  by the EU Commission through the project LASAGNE (FP7-ICT-318132).

\section*{References}
\begin{enumerate}
\item Levitt, B., March, J.G.
Organizational learning.
\emph{Ann Rev Socio} \textbf {14}: 319-340 (1988).

\item Delios, A., Henisz, W.J.
Political hazards, experience, and sequential entry strategies: the international expassion of Japanese firms 1980-1998.
\emph{ Int Manage J} \textbf {24}:1153-1164 (2003).

\item  Schlag, K.
Why imitate, and if so, how? A boundedlly rational approach to multi-armed bandits.
\emph{J Econ Theory} \textbf {78}:130-156 (1998).

 \item Vilone, D., Ramasco, J.J., S\'anchez, A., San Miguel, M.
Social and strategic imitation: the way to consensus.
\emph{Sci. Rep}. \textbf {2,} 686; DOI: 10.1038/srep00686 (2012).

\item Vega-Redondo, F.
Economics and the Theory of Games;
Cambridge University Press: Cambridge, UK, 2003.

\item Luthi, L., Pestelacci, E., Tomassini, M.
Cooperation and community structure in social networks.
\emph{Physica A} \textbf{ 387}, 955-966 (2008).

\item Skyrms, B.
The Stag Hunt and the Evolution of Social Structure;
Cambridge University Press: Cambridge, UK, 2004.

 \item Roca, C.P., Cuesta, J.A., S\'anchez, A.
Evolutionary Game theory: temporal and spatial effects beyond replicator dynamics.
\emph{Phys. Life Rev}. \textbf {6}, 208-249 (2009).

\item Tomassini, M., Pestelacci, E.
Coordination Games on Dynamical Networks.
\emph{Games} \textbf{1}, 242-261 (2010).

\item Scott, J.
Social Network Analysis. SAGE Publications, 2012.

\item Wasserman, S., Faust, K.
Social Network Analysis: Methods and Applications. Cambridge University Press, 1994.

\item Zhang, P., Peeta, S., Friesz, T.
Dynamic Game Theoretic Model of Multi-Layer Infrastructure Networks.
\emph{Netw Spat Econ} \textbf{ 5}, 147-178 (2005).

\item Kivel$\ddot{a}$, M., Arenas, A., Barthelemy, M., Gleeson, J.P., Moreno, Y., Porter, M.A.
Multilayer Networks	
arXiv:1309.7233 \emph{ J. Complex Netw.} \textbf{ 2}(3): 203-271 (2014).

\item Wang, Z. , Szolnoki, A., Perc, M.
Interdependent network reciprocity in evolutionary games.
\emph{Sci. Rep.} \textbf{3}, 1183 (2013).

\item Wang, Z., Szolnoki, A., Perc, M.
Optimal interdependence between networks for the evolution of cooperation.
\emph{Sci. Rep.} \textbf{3}, 2470 (2013).

\item Jiang, L.-L., Perc, M.
Spreading of cooperative behaviour across interdependent groups.
\emph{Sci. Rep}. \textbf{3}, 2483 (2013).

\item Szolnoki, A., Perc, M.
Information sharing promotes prosocial behaviour.
\emph{New J. Phys.} \textbf{15}, 053010 (2013).

\item Wang Z.,Wang L, Perc M.
Degree mixing in multilayer networks impedes the evolution of cooperation.
\emph{Phys. Rev. E} \textbf{89}, 052813 (2014).

\item Granovetter, M.
Threshold Models of Collective Behavior.
\emph{J. Am. Soc}. \textbf{83}, 1420-1443 (1978).

\item Centola, D., Egu\'iluz, V. M., Macy, M.W.
 Cascade Dynamics of Complex Propagation.
 \emph{Physica A} \textbf{374}, 449-456 (2007).

\item Gonz‡lez-Avella, J.C.,  Egu\'i'luz, V.M.,  Marsili, M., Vega-Redondo, F., San Miguel, M.
Threshold learning dynamics in social networks.
\emph{PLoS ONE}\textbf{ 6}(5), e20207 (2011).

\item Suchecki, K., Egu\'iluz, V. M., San Miguel, M.
Voter model dynamics in complex networks: Role of dimensionality.
\emph{Phys. Rev. E} \textbf{72}, 036132(1-8) (2005).

\item Ramsey FP Truth and probability in Ramsey, 1931. In Braithwaite RB (ed) The foundation of
mathematics and other logical essays, Chapter VII. Kegan, Paul, trench $\&$ Co., London; Harcourt, Brace and
Company, New York, 156-198. (1926).

\item Cabrales, A., Uriarte, J.R.
Doubts and Equilibria.
\emph{J Evol Econ }\textbf{23}, 783-810 (2013).

\item Gracia-L‡zaro, C., Ferrera, A., Ruiz,  G., Taranc\'on, A.,  Cuesta, J.A., Moreno, Y.,  S\'anchez,
A. Heterogeneous networks do not promote cooperation when humans play a Prisoner's Dilemma,
\emph{ PNAS }(USA) \textbf{109}, 12922-12926 (2012)

 \item Gruji\'c, J., Fosco, C.,  Araujo, L., Cuesta, J.A., S\'anchez A.
Social experiments in the mesoscale: Humans playing a spatial Prisoner's Dilemma
\emph{PLoS ONE }\textbf{ 5}(11), e13749 (2010).

\item Gruji\'c, J., Gracia-L\'azaro, C.,  Traulsen, A., Milinski, M., Semmann, D.,  Cuesta J.A., Moreno, Y., S\'anchez A.
A meta-analysis of spatial Prisoner's Dilemma experiments: Conditional cooperation and payoff irrelevance
\emph{Sci.Rep.}\textbf{ 4}, 4615 (2014).

 \item Vilone, D., Ramasco, J.J., S\'anchez, A., San Miguel, M.
Social imitation vs strategic choice, or consensus vs cooperation in the networked PrisonerÕs Dilemma
\emph{Physical Review E} \textbf{90}, 022810 (2014).

\item Nowak, M.A.,May, R.M.
Evolutionary games and spatial chaos.
\emph{Nature}  \textbf{359}, 826-829. (1992).

\item Kandori, M., Mailath, G., Rob, R.
Learning, mutation, and long-run equilibria in games.
\emph{Econometrica }\textbf{ 61}, 29-56. (1993).

\item Ellison, G.
Learning, local interaction, and coordination.
\emph{Econometrica}\textbf{ 61}, 1047-1071 (1993).

\end{enumerate}

\end{document}